# Locally Resonant Granular Chain

Luca Bonanomi[1], Georgios Theocharis[2,3], and Chiara Daraio[1,2]

[1] *Department of Mechanical and Process Engineering (D-MAVT) ETH Zurich, Zurich, Switzerland*
[2] *Graduate Aerospace Laboratories (GALCIT) California Institute of Technology, Pasadena, CA 91125, USA*
[3] *LAUM, CNRS, Université du Maine, Avenue O. Messiaen, 72085 Le Mans, France*

We report the design and testing of a tunable and nonlinear mechanical metamaterial, called locally resonant granular chain. It consists of a one-dimensional array of hollow spherical particles in contact, containing local resonators. The resonant particles are made of an aluminium outer spherical shell and a steel inner mass connected by a polymeric plastic structure acting as a spring. We characterize the linear spectra of the individual particles and of one-dimensional arrays of particles using theory, numerical analysis, and experiments. A wide band gap is observed as well as tunability of the dispersive spectrum by changing the applied static load. Finally, we experimentally explore the nonlinear dynamics of the resonant particles. By using nonlinear acoustical techniques, we reveal a complex, nonclassical nonlinear dynamics.

Artificial materials with a designed geometrical structure and selected material properties have been shown to present unusual behaviours with respect to electromagnetic [1] and acoustic wave propagation [2]. These media, called metamaterials, give rise to many novel phenomena, including subwavelength focusing and cloaking [3,4]. They also set the basis for the design of next-generation devices with tunable and switchable functionalities [5]. Many of these functionalities can be realized due to nonlinear processes emerging in the response of these structured materials when excited by external stimuli.

In acoustics, locally resonant acoustic metamaterials (AMs) [6] derive their unique properties from local resonators contained within each unit cell of a periodic structure. Typical examples are periodic arrangements of coated spheres/cylinders embedded in a linear homogeneous host medium [7] and arrays of Helmholtz resonators [8]. In contrast with their electromagnetic counterparts, for which there are numerous studies on their nonlinear dynamical behaviour, the research on the acoustics metamaterials has been mostly limited to the linear dynamical behaviour.

A great amount of work has been devoted to the study of *granular chains*, which are closely packed, ordered arrangements of elastic particles (spheres in most of the cases) in contact. Due to the Hertzian contact interaction between the particles [9], the dynamic vibrational response of these structures can be nonlinear and tunable [10]. This makes granular chains a perfect example for the study of fundamental phenomena [11] and engineering applications including tunable vibration filters [12], acoustic lenses [13], and acoustic rectifiers [14].

In this work, we combine the field of granular chains with that of locally resonant AMs. We thus combine the concepts of propagation of elastic waves through tunable nonlinear contacts and local resonators to design nonlinear tunable mechanical metamaterials. We call these structures *locally resonant granular chains*. Similar locally resonant granular chains with external resonant masses where described earlier [15]. We focus on the tunable and nonlinear behaviour of granular particles that include local internal resonators (or mass-in-mass units, i.e., MinM). This new design, due to its geometry, is more suitable to be embedded in a matrix, for example as reinforcement in a composite material. The MinM particles (Fig. 1a) are made of an outer mass (spherical shell), an inner mass (solid sphere) and an elastic structure between these two. In particular, the outer mass is made of an aluminum hollow shell (alloy 3003, diameter = 18.6 mm, wall thickness = $0.8 \pm 0.1$ mm, mass $m_1 = 2.3 \pm 0.05$ g, Young's modulus = 69 GPa, Poisson's ratio = 0.33). The inner mass consists of a solid steel bead (316 type, diameter = 12.8 mm, $m_2 = 8.6 \pm 0.1$ g, Young's modulus = 193 GPa, Poisson's ratio = 0.3). The reported values of Young's moduli and Poisson's ratios are standard specifications [16]. The elastic structure connecting the two masses is made of a commercial plastic (*Verowhite Plus*, by Objet, Young's modulus 2500 MPa [17]), has a mass of 0.2 g, and it has been created through 3D printing.

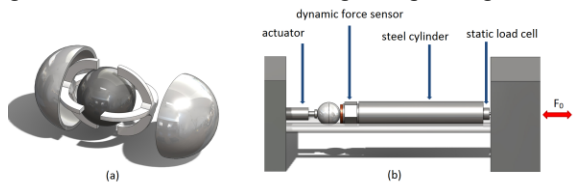

**Figure 1** (Color Online). (a) Parts and assembly of a MinM unit cell. (b) Schematic of the experimental setup to test the dynamic response of a single particle.

To test the dynamic response of a single particle and of a chain, we align and hold the particle(s) in a support system consisting of four polycarbonate rods, held in place by polycarbonate guide plates [12]. An initial static compression is applied on one side by using a translating steel cube. The static force applied is measured with a piezoelectric static load cell (Fig. 1b). We continuously drive the particle(s) with a piezo-electric actuator mounted on a fixed steel cube, on the opposite side with respect to the static force sensor. We record the transmitted force-time history using a dynamic force sensor. A heavy steel cylinder is added to the system in order to measure simultaneously the dynamic force and the static load.

The nonlinear interaction law between two elastic particles results from the Hertzian contact [9]. The Hertzian contact relates the contact force $F_{i,i+1}$ between two particles ($i$ and $i+1$) to the relative displacement $\Delta_{i,i+1}$ of their particle centers, as shown in the following equation:

$$F_{i,i+1} = A_{i,i+1}[\Delta_{i,i+1}]_+^{n_{i,i+1}} \quad (1).$$

Values inside the bracket $[s]_+$ only take positive values, which denotes the tensionless characteristic of the system (i.e., there is no force between the particles when they are separated). For two spheres (or a sphere and a flat surface; $R \rightarrow \infty$):

$$A_{i,i+1} = \frac{4E_i E_{i+1}\sqrt{\frac{R_i R_{i+1}}{R_i + R_{i+1}}}}{3E_{i+1}(1-v_i^2)+3E_i(1-v_{i+1}^2)}, \quad n_{i,i+1} = \frac{3}{2},$$

where $E_i$, $v_i$, $R_i$ are the elastic modulus, the Poisson's ratio, and the radius of the i-th particle, respectively. The exponent $n_{i,i+1} = 3/2$ comes from the geometry of the contact between two linearly elastic particles with elliptical contact area [9]. The contact interaction between thin hollow spheres was studied in [18]. For the range of wave amplitude considered in [18], a power-law type relation ($F=k\delta^n$) was used. It was found that the exponent $n$ was smaller than the value 3/2 as in the classical Hertzian interaction between solid spheres. The contact stiffness $k$ and the exponent $n$ were also found to be dependent on the thickness of the hollow sphere's shell. This dependence of the dynamic behavior of granular chains on the shell thickness of spherical particles provides yet another free parameter to employ in tuning the dynamics of nonlinear acoustic chains.

To describe the dynamics of the locally resonant granular chains considered in this work, we adopt a lumped-element approach. The validity of this approach is based on the fact that for the applied frequency range (0-20 kHz), collective resonant vibrations of the individual parts of the MinM unit cell are not excited (in order to verify this we computed the vibrational analysis of each part of a MinM finding that every resonant mode is above 30 kHz). Thus, the system can be considered as a mass-in-mass lattice [19] and its linear response is described by the following equations of motion:

$$m_1 \ddot{u}_1^{(j)} = k_1\left(u_1^{(j-1)} - 2u_1^{(j)} + u_1^{(j+1)}\right) + k_2\left(u_2^{(j)} - u_1^{(j)}\right) + \eta\left(\dot{u}_2^{(j)} - \dot{u}_1^{(j)}\right)$$
$$m_2 \ddot{u}_2^{(j)} = k_2\left(u_1^{(j)} - u_2^{(j)}\right) + \eta\left(\dot{u}_1^{(j)} - \dot{u}_2^{(j)}\right), \quad (2)$$

where $u_{1,2}^{(j)}$ represents the displacement of mass "$1,2$" in the $j^{th}$ cell. The elastic structure, being much softer of the two masses, plays the role of a linear damped spring with an effective stiffness $k_2$ and a viscous damping coefficient $\eta$. Finally, $k_1$ describes the linearized contact stiffness and depends on the static precompression $F_0$ as $k_1 = nA^{\frac{1}{n}}F_0^{\frac{n-1}{n}}$.

First, we studied the acoustic response of a single MinM particle (Fig. 2).

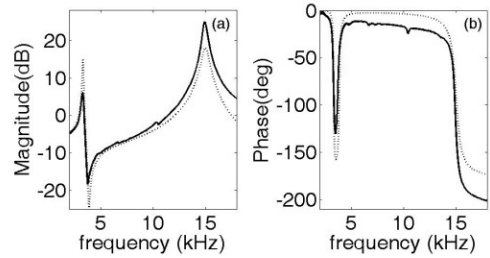

**Figure 2.** Experimental (solid line) and numerical (dashed line) magnitude (a) and phase (b) of the transfer function of a single MinM particle in the linear regime.

The characteristic spectrum of a single MinM compressed between two fixed walls (representing the actuator and the dynamic force sensor) presents two peaks, corresponding to the in-phase (at lower frequency) and to the out-of-phase (at higher frequency) motion of the two masses. Ignoring the dissipation, these eigenfrequencies are given by:

$$f_{1,2} = \frac{1}{2\pi}\sqrt{\frac{2k_1 m_2 + k_2(m_1+m_2) \pm \sqrt{-8k_1 k_2 m_1 m_2 + [2k_1 m_2 + (m_1+m_2)k_2]^2}}{2m_1 m_2}} \quad (3)$$

We characterize experimentally the acoustic properties of a single MinM exciting it with a low amplitude (approximately 10 mN peak) swept sine signal from 0.5 to 20 kHz. To analyze the data we compute the power spectral density (PSD) of the signal collected by the dynamic force sensor.

From Eq. (3) and the experimental vibrational peaks of a single MinM under a static load of $F_0$=17.3N (see Fig. 2a) we extract the following parameters, $k_2$ = 5.1·$10^6$ N/m and $k_1$=7.5·$10^6$ N/m. The latter corresponds to the linearized contact stiffness between the aluminum spherical shell and a stainless flat surface (actuator and force sensor). Considering a power law contact interaction ($F=A\delta^n$) and various static loads, we found that n=1.154 and A=4.583·$10^7$. Finally, we estimated the dissipation coefficient, $\eta$ =

8.6 kg/s, measuring the quality factor (and damping coefficient τ) of the out-of-phase resonant peak.

Fig 2 (a),(b) shows a very good agreement between the experimental and the numerical transfer function obtained by the state-space approach [20]. This proofs the validity of the lumped-element approach. A better agreement could be obtained using a more sophisticated modeling of the viscoelastic dissipation, but this is out of the scope of the present work. The transmission dip at $f_0$=3.8 kHz corresponds to an anti-resonance around which the dynamic mass density displays a resonance-like behavior (see chapter 5 of [2]). At that frequency, the displacement amplitude of the outer mass $m_2$, becomes zero as a result of a destructive interference of oscillations from the driver and the $m_1$ oscillator (the oscillations of the driver and the $m_1$ are out of phase). This is confirmed by the fact that $f_0 = 1/(2\pi)(k_2/m_2)^{1/2}$.

From the experimental acoustic response of two MinM unit cells, we derive the parameters for the fundamental contact law between the spherical shells ($F=A\delta^n$). We find that $n = 1.245$ and $A = 7.4 \cdot 10^7$ N/m$^{(n)}$. These values are in good agreement with those obtained with Finite Element Methods (FEM) in [18].

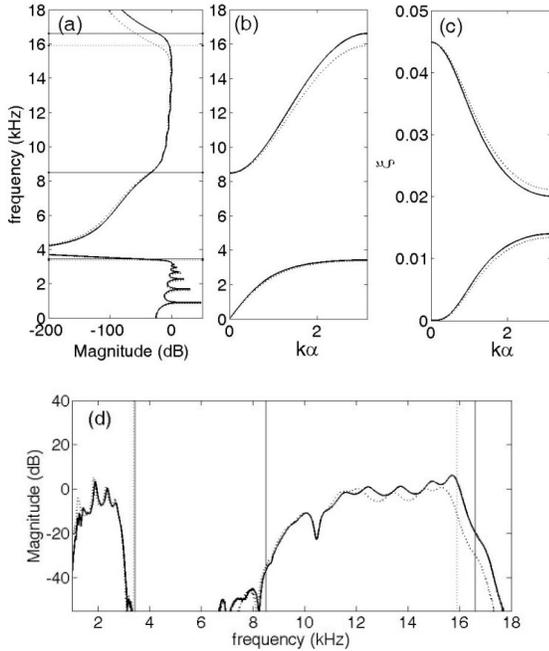

**Figure 3.** (a) Magnitude of the transfer function, (b) damped dispersion relation for an infinite 1D locally resonant granular chain using the experimentally obtained parameters and the (c) damping ratio band structure for $F_0$=22.6 N (solid lines) and $F_0$=12.6 N (dotted line) respectively. (d) Experimental transfer function of a chain of 11 MinMs under the applied load of 22.6 N (solid line) and 12.6 N (dotted line) respectively. The horizontal (vertical) lines in panel (a) (in panel (d)) denote the analytical values of the frequencies of the band edges.

In Fig. 3 (a), we show the numerical transfer function for a chain consisting of 11 units with two different initial static loads. The horizontal lines denote the following analytical values of the band edges obtained for the loss-less case ($\eta$=0):

$$f_1 = 1/(2\pi) \, [(b-(b^2-4ac)^{1/2})/2a)]^{1/2}$$

$$f_2 = 1/(2\pi) \, [k_2(m_1+m_2)/(m_1 m_2)]^{1/2}$$

$$f_3 = 1/(2\pi) \, [(b+(b^2-4ac)^{1/2})/2a)]^{1/2} \quad ,$$

where $a = m_1 m_2$, $b = k_2(m_1+m_2)+4m_2 k_1$ and $c = 4k_1 k_2$. The upper acoustic edge $\omega_1$ and the upper optical edge $\omega_3$ depend on $k_1$ and thus on the static load. The tunability of the contact stiffness $k_1$ upon changes to the static precompression $F_0$ is an important feature of the locally resonant granular chains.

Using the Bloch state-space formulation (see chapter 6 of [2]) and the experimentally obtained parameters, we derive the damped frequency band structure, Fig. 3(b), and the corresponding wavenumber-dependent damping ratio, plotted in Fig 3(c). From Fig 3 (b), it is clear the existence of a band gap. The origin of this gap is connected with hybridization effect of the usual dispersion curve of a monoatomic chain with local resonances. The lower (upper) pass bands consist of modes where the inner and outer masses are oscillating in phase (out of phase). From Fig 3 (c), we can conclude that the damping ratio of the optical branch modes is higher than that of the acoustic modes since the presence of the damped material influences mostly the out of phase motion of the two masses. Among the optical modes, the ones with lower wavenumber experience higher drop rates. Comparing the analytical expressions of the band edges and the damped frequency band structure, we can conclude that the presence of the dissipation has no influence on the dispersive properties of the structure.

We experimentally characterize the linear behavior of a locally resonant granular chain consisting of 11 resonant particles. As in the case of a single MinM, we continuously drive the chain applying low-amplitude (approximately 10 mN peak) swept sine signal from 0.5 to 20 kHz. We note that, for a static load of 22.6 N, a wide band gap exists between $f_1$=3.5 kHz and $f_2$=8 kHz as well as a gap above $f_3$=17 kHz. A good agreement between experiments, Fig. 3(d), and theory, Fig. 3(a) is evident. It is possible to tune the band frequencies by varying the static precompression. To demonstrate this, we decreased the static preload from 22.6 N to 12.6 N. In this case, the upper cut-off frequency of the optical band downshifts of about 1 kHz, while the upper acoustic edge and the lower optical edge shift only a few Hz and their tunability is not visible in experiments. The dip around 11 kHz, which is not captured by the theoretical model, comes from the cavity resonance of the air contained within the MinM [21].

We also explore the nonlinear dynamics of our structure triggered by increasing the amplitude of the excitation. In Fig. (4), one can see the nonlinear resonance response of a single MinM particle for different amplitudes of the input signal. Fig. 4(a) is the first resonance (the one corresponding to the in-phase motion of the two masses); Figure 4(b) is the second resonance (the one corresponding to the out-of-phase motion of the two masses). Experiments clearly show that a softening effect emerges for both resonances, when the driving amplitude is increased. In particular, the resonance frequency shift $\Delta f$ of the first mode, Fig. 4(c), follows a quadratic softening as a function of the drive amplitude. This is in agreement with the prediction of a simple phenomenological model (Duffing) [22] and in accordance with weakly nonlinear Hertzian nonlinearity [23].

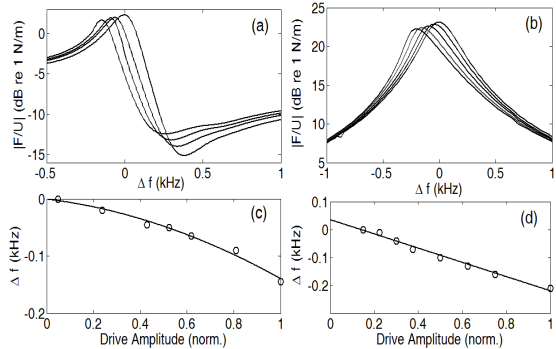

**Figure 4.** (a),(b) Experimentally measured acoustic transfer function (force from dynamic sensor over actuator's displacement) and (c),(d) resonance frequency shift as a function of the drive amplitude around the in-phase motion resonance (a),(c) and the out-of-phase motion resonance peak (b),(d).

However, the second resonance, which corresponds to the out-of-phase motion, seems to follow a linear dependence on the amplitude. This linear amplitude dependence is commonly observed in a wide class of materials called mesoscopic media [24], including damaged solids and geomaterials, and is attributed to hysteretic quadratic nonlinearity. In the case of granular media, it has been shown that frictional nonlinearities (for example the single shear contact between two spheres) exhibits nonlinear hysteresis [25]. In our case, we believe that this nonclassical nonlinear response of the second mode is induced by the presence of the soft elastic layer. Nonlinear hysteretic mechanisms originate from the complex, frictional contact forces between the elastic structure with the two masses, and other mechanisms such as thermoelastic or viscous losses [26], could explain the linear amplitude-dependent frequency downshift. These mechanisms could also explain the weak, amplitude-dependent variation in the dissipation that mostly the second resonance exhibit.

*Conclusions.* We have designed and built a novel acoustic periodic structure composed of granular particles enclosing mechanical resonators. The local resonances placed along the chain we assembled, resulted in the existence of a wide band gap in the audible regime that makes our system capable of filtering mechanical waves between 3 and 8.5 kHz. We have characterized the dynamics of the compressed 1D chain using theory, numerical simulations and experiments. We found good agreement for the linear spectrum and we explored experimentally the nonlinear behaviour of the system applying nonlinear resonant spectroscopy. The presence of the elastic structure leads to a nonclassical nonlinear dynamics. Our experimental results are a first step towards using nonlinearity as a way to tune the dynamic and acoustic properties of phononic crystals and acoustic metamaterials. Due to their geometry, MinM particles can easily be embedded in a matrix and, for this reason, our work can represent an important step in coupling vibration mitigation and reinforcement in composite materials.

*Acknowledgments* – L.B. and C.D. acknowledge support from the US NSF, grant number 0969541 and the US Office of Naval Research, YIP program. G.T. acknowledges support from CIG FP7 ComGranSol. We thank Marc Serra Garcia for his support in experiments. We acknowledge V. Tournat and D. Ngo for helpful discussions.